\documentclass{article}
\usepackage{spconf,amsmath,graphicx,hyperref}
\usepackage[capitalize]{cleveref}
\usepackage{mathtools}
\usepackage{amssymb}
\usepackage{multirow}
\usepackage{booktabs}
\usepackage{xcolor}
\usepackage{adjustbox}

\crefname{section}{Sec.}{Secs.}
\Crefname{section}{Section}{Sections}
\Crefname{table}{Table}{Tables}
\crefname{table}{Tab.}{Tabs.}

\title{MotionBeat: Motion-Aligned Music Representation via Embodied Contrastive Learning and Bar-Equivariant Contact-Aware Encoding}
\name{Xuanchen Wang \qquad Heng Wang \qquad Weidong Cai}
\address{School of Computer Science, The University of Sydney, Australia
}
\begin{document}
%
\maketitle
\begin{abstract}
Music is both an auditory and an embodied phenomenon, closely linked to human motion and naturally expressed through dance. However, most existing audio representations neglect this embodied dimension, limiting their ability to capture rhythmic and structural cues that drive movement. We propose MotionBeat, a framework for motion-aligned music representation learning. MotionBeat is trained with two newly proposed objectives: the Embodied Contrastive Loss (ECL), an enhanced InfoNCE formulation with tempo-aware and beat-jitter negatives to achieve fine-grained rhythmic discrimination, and the Structural Rhythm Alignment Loss (SRAL), which ensures rhythm consistency by aligning music accents with corresponding motion events. Architecturally, MotionBeat introduces bar-equivariant phase rotations to capture cyclic rhythmic patterns and contact-guided attention to emphasize motion events synchronized with musical accents. Experiments show that MotionBeat outperforms state-of-the-art audio encoders in music-to-dance generation and transfers effectively to beat tracking, music tagging, genre and instrument classification, emotion recognition, and audio–visual retrieval. Our project demo page: \href{https://motionbeat2025.github.io/}{https://motionbeat2025.github.io/}.
\end{abstract}
\section{Introduction}
\label{sec:intro}
Music is inherently embodied: people instinctively synchronize their movements with rhythm, linking sound and motion in ways central to dance and rhythm understanding. However, existing audio representation models overlook this embodied dimension, relying on audio–text \cite{elizalde2023clap, wu2023large, elizalde2024natural} or audio–visual \cite{wu2022wav2clip, girdhar2023imagebind} training that captures semantics and acoustics but not movement. This gap leaves current audio encoders prone to rhythm–motion misalignment and limits their effectiveness in music-to-dance generation \cite{tseng2023edge, li2021ai, wang2025choreomuse, wang2025dance} and related tasks.

To address this gap, we propose MotionBeat, a framework for motion-aligned music representation learning. MotionBeat is designed to learn music embeddings that are directly grounded in human motion, capturing not only auditory features but also the embodied rhythmic structures that govern dance. To achieve this, we introduce two novel objectives: Embodied Contrastive Loss (ECL), which enhances InfoNCE \cite{oord2018representation} with tempo-aware and beat-jitter negatives for fine-grained rhythmic discrimination, and Structural Rhythm Alignment Loss (SRAL), which enforces beat- and bar-level consistency via Soft-DTW \cite{cuturi2017soft} alignment between audio onsets and motion contacts, and optimal transport alignment between accent mass and motion energy.

From an architectural perspective, our MotionBeat relies on two fundamental mechanisms tailored for embodied music representation. The first, bar-equivariant phase rotations, encode the cyclic nature of rhythm by applying rotational transformations to embeddings. This enforces structural consistency under phase shifts and makes representations robust to different bar starting points. The second, contact-guided attention, highlights motion frames where embodied events occur by weighting attention scores with contact probabilities. This directs more representational capacity to beats aligned with musical accents, strengthening the audio–motion coupling.

We evaluate MotionBeat across both generative and recognition tasks. In music-to-dance generation, it surpasses state-of-the-art audio encoders, producing motion that is more rhythmically aligned and structurally consistent. Moreover, MotionBeat transfers effectively to downstream tasks such as beat tracking, music tagging, genre and instrument classification, emotion recognition, and audio–visual retrieval.

Our contributions can be summarized as follows: 
\begin{itemize}
\item {}We introduce MotionBeat, a framework for motion-aligned music representation learning.

\item {}We propose two new training objectives, Embodied Contrastive Loss (ECL) and Structural Rhythm Alignment Loss (SRAL), to jointly capture fine-grained rhythmic detail and global structural consistency.

\item {}We develop architectural innovations, including bar-equivariant phase rotations and contact-guided attention, that explicitly model cyclic rhythm and emphasize motion-synchronized events.

\item {}Extensive experiments show that MotionBeat surpasses state-of-the-art encoders in dance generation and generalizes to beat tracking, tagging, classification, emotion recognition, and cross-modal retrieval.

\end{itemize}

\section{Method}

\subsection{Training Objective}
\label{subsec:loss}
Standard contrastive learning usually relies on random in-batch negatives. In music–motion alignment, these are too easy since negatives often differ in genre, timbre, or instrumentation, letting the model rely on global acoustic cues while ignoring rhythmic alignment.

\subsubsection{Embodied Contrastive Loss}
Embodied Contrastive Loss (ECL) extends the standard InfoNCE \cite{oord2018representation} objective by introducing rhythm-sensitive negatives. Tempo-aware negatives share a similar BPM but differ in bar-phase or accent patterns, forcing the model to look beyond global tempo. Beat-jitter negatives are generated by shifting motion or audio by $\pm 1$ beat within the same clip, keeping acoustics and style intact but disrupting step timing. Given a batch of $N$ paired music--motion clips $\{(\mathbf{z}_a^i, \mathbf{z}_m^i)\}_{i=1}^N$, 
each anchor $\mathbf{z}_a^i$ has one positive (its paired motion embedding $\mathbf{z}_m^i$) and multiple categories of negatives. The denominator $D^i$ is defined as:  
\begin{align}
D^i &= \exp \left( s(\mathbf{z}_a^i, \mathbf{z}_m^i) / \tau \right) \notag \\
&+ \sum_{j \in \mathcal{N}_{\text{batch}}} \exp \left( s(\mathbf{z}_a^i, \mathbf{z}_m^j) / \tau \right) \notag \\
&+ \sum_{k \in \mathcal{N}_{\text{tempo}}} \exp \left( s(\mathbf{z}_a^i, \mathbf{z}_m^k) / \tau \right) \notag \\
&+ \sum_{l \in \mathcal{N}_{\text{jitter}}} \exp \left( s(\mathbf{z}_a^i, \mathbf{z}_m^l) / \tau \right),
\end{align}
where $\tau$ is a temperature parameter, $s(\cdot,\cdot)$ denotes cosine similarity, $\mathcal{N}_{\text{batch}}$ are random in-batch negatives,  
$\mathcal{N}_{\text{tempo}}$ are tempo-aware negatives, and $\mathcal{N}_{\text{jitter}}$ are beat-jitter negatives.  

The ECL objective is then defined as:  
\begin{equation}
\mathcal{L}_{\text{ECL}} = - \frac{1}{N} \sum_{i=1}^{N} 
\log \frac{\exp \left( s(\mathbf{z}_a^i, \mathbf{z}_m^i) / \tau \right)}{D^i}.
\end{equation}

\subsubsection{Structural Rhythm Alignment Loss}

While ECL enforces pairwise discriminability, it does not explicitly model higher-level rhythmic structure. We therefore introduce the Structural Rhythm Alignment Loss (SRAL), which aligns audio and motion at both the beat and bar levels.

For beat-level alignment, let $\mathbf{o}_{1:K}$ denote the audio onset envelope across $K$ beats, and $\mathbf{c}_{1:K}$ the motion contact pulse sequence. We apply differentiable dynamic time warping (Soft-DTW) \cite{cuturi2017soft}:  
\begin{equation}
\mathcal{L}_{\text{beat}} = \text{SoftDTW}(\mathbf{o}_{1:K}, \mathbf{c}_{1:K}),
\end{equation}
which tolerates small deviations while rewarding synchrony at the beat level.  

At the bar level, we treat audio as a distribution of accent mass $\mathbf{a}_{\text{bar}}$ and motion as a distribution of kinetic energy $\mathbf{m}_{\text{bar}}$, both normalized within each bar.  
We align them using Earth Mover’s Distance (EMD) \cite{rubner2000earth}:  
\begin{equation}
\mathcal{L}_{\text{bar}} = \text{EMD}(\mathbf{a}_{\text{bar}}, \mathbf{m}_{\text{bar}}),
\end{equation}
which measures the minimal effort to transform the audio accent distribution into the motion energy distribution and thus naturally captures flexible bar-level rhythmic shifts.

The combined SRAL is:  
\begin{equation}
\mathcal{L}_{\text{SRAL}} = \lambda_{\text{beat}} \, \mathcal{L}_{\text{beat}} + \lambda_{\text{bar}} \, \mathcal{L}_{\text{bar}},
\end{equation}
where $\lambda_{\text{beat}}$ and $\lambda_{\text{bar}}$ balance beat- and bar-level terms, set to 0.9 and 0.2, respectively.

\subsubsection{Overall Loss}
Finally, the total training objective combines ECL and SRAL:  
\begin{equation}
\mathcal{L}_{\text{total}} = \mathcal{L}_{\text{ECL}} + \alpha \, \mathcal{L}_{\text{SRAL}},
\end{equation}
where $\alpha$ controls the trade-off between contrastive alignment and structural rhythm alignment, and is empirically set to 0.2 in our experiments.

\begin{figure}[t]
  \centering
  \includegraphics[width=\linewidth]{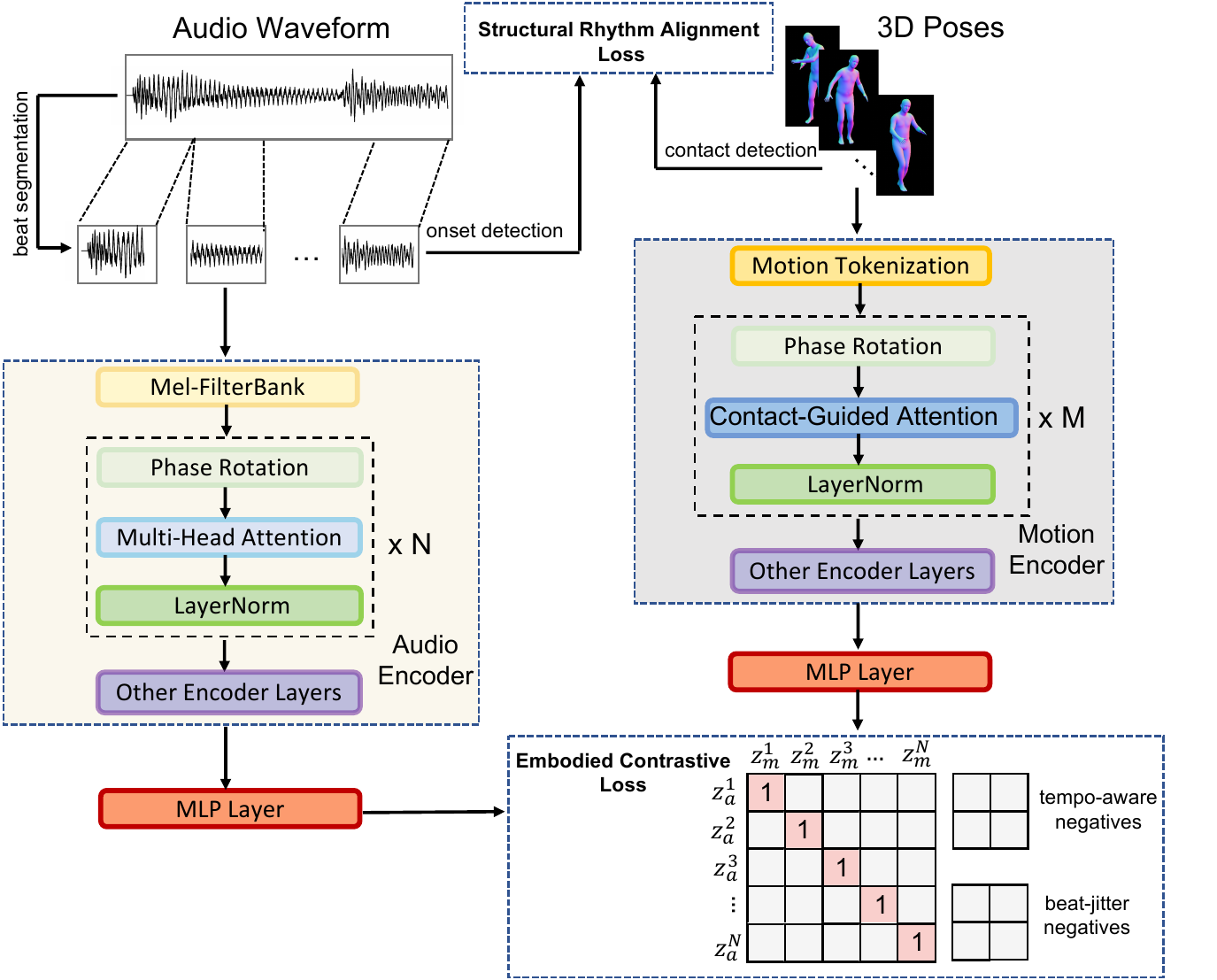}
  \caption{Architecture of MotionBeat. Audio and motion inputs are converted into beat-synchronous tokens and passed through encoders with bar-equivariant phase rotations and contact-guided attention. Embeddings are trained with the Embodied Contrastive Loss (ECL), while auxiliary rhythm heads provide onset envelopes and contact pulses for the Structural Rhythm Alignment Loss (SRAL). The total loss integrates both objectives to learn motion-aligned music representations.}
  \label{fig:model}
\end{figure}

\subsection{Model Architecture}
MotionBeat, illustrated in  \cref{fig:model}, consists of two modality-specific encoders and a lightweight set of task heads that feed the losses in \cref{subsec:loss}. The audio encoder $f_a$ maps each beat-synchronous audio segment to an embedding $\mathbf{z}_a \in \mathbb{R}^{d}$ using $N$ blocks that integrate phase rotation, self-attention, and layer normalization. Similarly, the motion encoder $f_m$ maps temporally aligned motion segments to embeddings $\mathbf{z}_m \in \mathbb{R}^{d}$ through $M$ blocks combining phase rotation, contact-guided attention, and layer normalization. Two projection heads then map the encoder states into a shared contrastive space for ECL. In parallel, we detect audio onsets and motion contacts to compute onset envelopes, bar accent mass, contact pulses, and bar energy mass, which serve as inputs to SRAL.

\subsubsection{Input Representation}
All inputs are represented in a beat-synchronous form.  
Rather than processing at the raw frame level, we segment both audio and motion into $K$ consecutive beat intervals, using estimated tempo and downbeat information. For the audio input, we extract a log-mel spectrogram for each beat interval and apply average pooling across it. This yields one audio token $\mathbf{x}^a_t \in \mathbb{R}^F$ per beat, where $t=1,\dots,K$. Bar-phase encodings are added as extra positional features. For motion, we start from 3D body joints and SMPL \cite{loper2023smpl} parameters. We compute per-frame kinematic features, then average-pool them within each beat interval. This produces one motion token $\mathbf{x}^m_t \in \mathbb{R}^M$ per beat, aligned with the corresponding audio token.




\subsubsection{Bar-Equivariant Phase Rotations}

Within a bar of $B$ beats, we want embeddings to transform equivariantly under cyclic phase shifts. Let $\phi_t = 2\pi (t \bmod B)/B$ denote the bar phase of token $t$. For each attention head, we pair channels into 2D planes and apply a complex rotation to queries and keys before the dot-product:
\begin{equation}
\tilde{\mathbf{q}}_t = \mathcal{R}(\phi_t)\,\mathbf{q}_t,\qquad
\tilde{\mathbf{k}}_u = \mathcal{R}(\phi_u)\,\mathbf{k}_u,
\end{equation}
where $\mathcal{R}(\phi)$ applies $(\cos\phi, -\sin\phi; \sin\phi, \cos\phi)$ to each 2D channel pair. A phase shift by $\Delta$ beats ($\Delta\phi = 2\pi\Delta/B$) induces the same rotation on all tokens, so a cyclic shift in time corresponds to a rotation in latent space. This enforces within-bar consistency and improves generalization to different bar starting points.

\begin{table*}[t]
\centering
\begin{adjustbox}{width=\textwidth}
\normalsize   
\begin{tabular}{c|ccccccccccccc}
    \toprule
    \multirow{2}{*}{Method} & \multicolumn{4}{c}{Dance Generation} & \multicolumn{3}{c}{Beat Tracking} & \multicolumn{2}{c}{Music Tagging} & Genre Classification & Instrument Classification & \multicolumn{2}{c}{Emotion Recognition} \\
    \cmidrule(lr){2-5} \cmidrule(lr){6-8} \cmidrule(lr){9-10} \cmidrule(lr){11-12} \cmidrule(lr){13-14} 
              & PFC $\downarrow$ & $\text{Dist}_k$ $\uparrow$ & $\text{Dist}_g$ $\uparrow$ & BAS $\uparrow$ & F1 $\uparrow$ & CML\textsubscript{t} $\uparrow$ & AML\textsubscript{t} $\uparrow$ & ROC $\uparrow$ & AP $\uparrow$ & ACC $\uparrow$ & ACC $\uparrow$ & R2\textsuperscript{V} $\uparrow$ & R2\textsuperscript{A} $\uparrow$ \\ 
    \midrule
    wav2vec 2.0 \cite{baevski2020wav2vec} & 1.698 & 8.21 & 5.56 & 0.23 & 0.845 & 0.791 & 0.878 & 89.1 & 39.2 & 68.6 & 66.7 & 45.2 & 66.5 \\
    CLAP \cite{wu2023large} & 1.625 & 9.53 & 5.75 & 0.25 & 0.851 & 0.795 & 0.885 & 88.8 & 39.1 & 69.3 & 65.8 & 48.8 & 69.8 \\
    Wav2CLIP \cite{wu2022wav2clip} & 1.602 & 10.65 & 6.39 & 0.24 & 0.848 & 0.801 & 0.879 & 89.5 & 39.1 & 71.2 & 67.2 & 50.3 & 70.1 \\
    Jukebox \cite{dhariwal2020jukebox} & 1.598 & 10.75 & 7.03 & 0.24 & 0.865 & \textbf{0.804} & 0.892 & 90.5 & 40.3 & 78.5 & 68.8 & \textbf{61.5} & 72.1 \\
    \midrule
    MotionBeat (w/o ECL) & 1.588 & 10.88 & 7.55 & 0.25 & 0.858 & 0.801 & 0.886 & 90.4 & 39.8 & 77.9 & 68.9 & 59.3 & 71.5 \\
    MotionBeat (w/o SRAL) & 1.590 & 10.21 & 7.65 & 0.24 & 0.835 & 0.799 & 0.879 & 89.8 & 40.2 & 78.1 & 68.2 & 59.8 & 72.3 \\
    MotionBeat & \textbf{1.545} & \textbf{11.02} & \textbf{7.89} & \textbf{0.27} & \textbf{0.878} & 0.803 & \textbf{0.905} & \textbf{91.2} & \textbf{40.8} & \textbf{79.2} & \textbf{70.8} & 61.2 & \textbf{73.8} \\
    \bottomrule
\end{tabular}
\end{adjustbox}
\caption{Comparison of different audio encoders and MotionBeat (with ablations) across dance generation and transfer tasks. Variants “w/o ECL” and “w/o SRAL” denote ablations without the Embodied Contrastive Loss and Structural Rhythm Alignment Loss, respectively.}
\label{tab:dance comparison}
\end{table*}

\begin{table}[t]
\centering
\begin{adjustbox}{width=\linewidth}
\normalsize   
\setlength{\tabcolsep}{4pt}
\begin{tabular}{ccccc|cccc}
\toprule
\multirow{2}{*}{Method} & \multicolumn{4}{c}{Music $\rightarrow$ Motion} & \multicolumn{4}{c}{Motion $\rightarrow$ Music} \\
\cmidrule(lr){2-5}\cmidrule(lr){6-9}
 & R@1 $\uparrow$ & R@5 $\uparrow$ & R@10 $\uparrow$ & MedR $\downarrow$ & R@1 $\uparrow$ & R@5 $\uparrow$ & R@10 $\uparrow$ & MedR $\downarrow$ \\
\midrule
wav2vec 2.0 \cite{baevski2020wav2vec} & 13.7 & 31.2 & 44.5 & 39 & 13.1 & 30.8 & 43.8 & 40 \\
CLAP \cite{wu2023large}               & 15.6 & 33.9 & 47.2 & 36 & 15.2 & 33.1 & 46.7 & 37 \\
Wav2CLIP \cite{wu2022wav2clip}        & 16.2 & 34.7 & 48.6 & 34 & 15.8 & 34.0 & 47.9 & 35 \\
Jukebox \cite{dhariwal2020jukebox}    & 19.8 & 40.5 & 53.2 & 27 & 18.8 & 38.9 & 52.5 & 29 \\
MotionBeat                            & \textbf{22.1} & \textbf{42.5} & \textbf{58.3} & \textbf{22} & \textbf{21.7} & \textbf{41.9} & \textbf{57.6} & \textbf{23} \\
\bottomrule
\end{tabular}
\end{adjustbox}
\caption{Cross-modal retrieval between music and motion on AIST++. Retrieval is evaluated in both directions using Recall@K and Median Rank.}
\label{tab:retrieval}
\end{table}

\begin{table}[t]
\centering
\begin{adjustbox}{width=\linewidth}
\small   
\setlength{\tabcolsep}{6pt}
\begin{tabular}{lcccc}
\toprule
Model Variant & BAS $\uparrow$ & Beat F1 $\uparrow$ & PFC $\downarrow$ & R@1 (M $\rightarrow$ Mtn) $\uparrow$ \\
\midrule
Baseline (no BEP, no CGA) & 0.24 & 0.852 & 1.60 & 19.3 \\
\ + BEP only              & 0.26 & 0.866 & 1.57 & 20.8 \\
\ + CGA only              & 0.25 & 0.871 & 1.56 & 20.3 \\
\ + BEP \& CGA            & \textbf{0.27} & \textbf{0.878} & \textbf{1.55} & \textbf{22.1} \\
\bottomrule
\end{tabular}
\end{adjustbox}
\caption{Ablation of architectural components on key metrics: beat alignment (BAS), rhythmic accuracy (Beat F1), physical plausibility (PFC), and cross-modal retrieval (Music$\rightarrow$Motion R@1). BEP = Bar-Equivariant Phase Rotations; CGA = Contact-Guided Attention.}
\label{tab:bep_cga_ablation}
\end{table}

\subsubsection{Contact-Guided Attention}
Let $r_t \in [0,1]$ be a contact probability at token $t$. We bias attention toward tokens that coincide with contacts. For each query position $t$ and key position $u$, the attention weights are defined as:
\begin{equation}
A_{tu} = \frac{\exp \left( \langle \tilde{\mathbf{q}}_t, \tilde{\mathbf{k}}_u \rangle / \sqrt{d_h} + \alpha_{\text{logit}} r_u \right)}
{\sum_{u'} \exp \left( \langle \tilde{\mathbf{q}}_t, \tilde{\mathbf{k}}_{u'} \rangle / \sqrt{d_h} + \alpha_{\text{logit}} r_{u'} \right)},
\end{equation}
where $d_h$ is the dimensionality of each attention head and $\alpha_{\text{logit}} \geq 0$ is a learnable scalar controlling logit bias.  
In addition, value vectors are reweighted as:
\begin{equation}
\mathbf{v}_u \leftarrow (1 + \alpha_{\text{val}} r_u)\,\mathbf{v}_u,
\end{equation}
with learnable $\alpha_{\text{val}} \geq 0$. This allocates representational bandwidth to embodied anchors of rhythm without discarding non-contact frames. During training, $r_t$ can be predicted by a small contact head.

\section{Experiments and Results}
\subsection{Dataset and Implementation Details}
We use the AIST++ dataset~\cite{li2021ai} of paired music--dance recordings with 3D skeleton annotations, representing audio as beat-synchronous log-mel tokens (128 bands) and motion as normalized SMPL kinematics with contact cues. Models are implemented on a single A6000 GPU using 6-layer Transformers (512 hidden units, 8 heads, 128-dimensional embeddings) and trained with AdamW \cite{loshchilov2017decoupled} (lr $2\times10^{-4}$, batch size $64$, $\tau=0.07$) for up to 100 epochs with early stopping.

\subsection{Downstream Tasks}
We evaluate MotionBeat on dance generation as the primary task and on a broad set of recognition tasks. For dance generation, we assess choreography quality along three axes: physical plausibility via the Physical Foot Contact (PFC) score \cite{tseng2023edge}, diversity \cite{siyao2022bailando, li2021ai} using distributional spreads in kinetic (Dist$_k$) and geometric (Dist$_g$) feature spaces, and music--motion alignment with the Beat Alignment Score (BAS) \cite{siyao2022bailando}. Beat tracking is evaluated on the GTZAN \cite{tzanetakis2002musical} dataset using standard metrics \cite{davies2009evaluation}: F1, CML\textsubscript{t}, and AML\textsubscript{t}. For music tagging, we use the MagnaTagATune (MTT) \cite{law2009evaluation} dataset with ROC and AP as evaluation metrics. Genre classification is performed on the GTZAN dataset with accuracy (ACC) as the measure. Instrument classification is conducted on NSynth \cite{engel2017neural}, also using ACC. Finally, emotion regression is evaluated on the Emomusic \cite{soleymani20131000} dataset, using the coefficient of determination ($R^2$) for arousal and valence.

\subsection{Baseline Methods}
We benchmark MotionBeat against four representative audio encoders that cover different learning paradigms. wav2vec 2.0 \cite{baevski2020wav2vec} is a self-supervised model pretrained on large-scale speech and audio, providing general-purpose representations. CLAP \cite{wu2023large} aligns audio and text in a joint embedding space, excelling at semantic transfer. Wav2CLIP \cite{wu2022wav2clip} aligns audio with vision–language features from CLIP, offering multimodal grounding and cross-domain generality. Jukebox \cite{dhariwal2020jukebox} is trained on large-scale music data with a hierarchical VQ-VAE, providing music-specific representations that capture pitch, timbre, and long-term structure. For fairness, we use the same protocol across tasks. All baseline encoders are frozen, and only lightweight heads are trained: a small regressor for beat tracking and emotion recognition, a linear classifier for genre and instrument classification, and a multi-label head for music tagging. For cross-modal retrieval, we add a projection head to map embeddings into a shared space while keeping the backbone fixed. For dance generation, we use EDGE \cite{tseng2023edge} to generate dance. This ensures that performance differences reflect the quality of the embeddings rather than task-specific fine-tuning.

\subsection{Result Analysis}
Table~\ref{tab:dance comparison} compares strong audio encoders with our MotionBeat and its ablated variants across all evaluation tasks. For \textbf{dance generation}, MotionBeat delivers the most balanced gains, producing more physically plausible dances (lowest PFC), greater motion diversity, and stronger music–motion alignment. This shows that MotionBeat avoids overfitting to limited patterns and better captures the rhythmic structure needed for synchronized choreography. In \textbf{beat tracking}, MotionBeat achieves the best F1 and AML$_t$, while Jukebox is competitive on CML$_t$. This shows that although music-specific encoders capture some rhythmic cues, MotionBeat offers a more fine-grained beat-level representation. In \textbf{music tagging}, MotionBeat consistently outperforms all baselines, achieving the highest ROC and AP, which demonstrates that embodied rhythm information enhances semantic prediction beyond timbral or textual alignment alone. Performance gains extend to \textbf{genre classification} and \textbf{instrument classification}, showing that MotionBeat preserves timbral and categorical information while improving rhythmic awareness. In \textbf{emotion recognition}, MotionBeat achieves the best arousal score and matches Jukebox on valence, confirming that embodied rhythm also contributes to modeling affective dimensions of music. The ablation studies highlight the complementary contributions of our objectives: removing ECL reduces rhythmic discrimination, while removing SRAL lowers structural alignment. Table~\ref{tab:retrieval} further shows that MotionBeat achieves superior cross-modal retrieval in both directions, outperforming Jukebox and other baselines. Finally, Table~\ref{tab:bep_cga_ablation} confirms that Bar-Equivariant Phase Rotations and Contact-Guided Attention each provide improvements, and their combination yields the strongest overall performance. These results demonstrate that MotionBeat’s rhythm-aware design is crucial for capturing embodied music structure.

\section{Conclusion}
We present MotionBeat, a framework for motion-aligned music representation learning. 
Through novel objectives and rhythm-aware architectural designs, MotionBeat captures embodied rhythmic structure and consistently outperforms strong audio encoders on dance generation and a range of recognition tasks. This demonstrates the value of motion as a supervisory signal for learning music representations.

\bibliographystyle{IEEEbib}
\bibliography{main}

@inproceedings{tseng2023edge,
  title={Edge: Editable dance generation from music},
  author={Tseng, Jonathan and Castellon, Rodrigo and Liu, Karen},
  booktitle={CVPR},
  year={2023}
}

@inproceedings{wang2025dance,
  title={Dance any beat: Blending beats with visuals in dance video generation},
  author={Wang, Xuanchen and Wang, Heng and Liu, Dongnan and Cai, Weidong},
  booktitle={WACV},
  year={2025}
}

@inproceedings{li2021ai,
  title={AI choreographer: Music conditioned 3D dance generation with AIST++},
  author={Li, Ruilong and Yang, Shan and Ross, David A and Kanazawa, Angjoo},
  booktitle={ICCV},
  year={2021}
}

@inproceedings{elizalde2023clap,
  title={CLAP: Learning audio concepts from natural language supervision},
  author={Elizalde, Benjamin and Deshmukh, Soham and Al Ismail, Mahmoud and Wang, Huaming},
  booktitle={ICASSP},
  year={2023}
}

@inproceedings{wu2023large,
  title={Large-scale contrastive language-audio pretraining with feature fusion and keyword-to-caption augmentation},
  author={Wu, Yusong and Chen, Ke and Zhang, Tianyu and Hui, Yuchen and Berg-Kirkpatrick, Taylor and Dubnov, Shlomo},
  booktitle={ICASSP},
  year={2023}
}

@inproceedings{wu2022wav2clip,
  title={Wav2clip: Learning robust audio representations from CLIP},
  author={Wu, Ho-Hsiang and Seetharaman, Prem and Kumar, Kundan and Bello, Juan Pablo},
  booktitle={ICASSP},
  year={2022}
}

@inproceedings{girdhar2023imagebind,
  title={Imagebind: One embedding space to bind them all},
  author={Girdhar, Rohit and El-Nouby, Alaaeldin and Liu, Zhuang and Singh, Mannat and Alwala, Kalyan Vasudev and Joulin, Armand and Misra, Ishan},
  booktitle={CVPR},
  year={2023}
}

@inproceedings{elizalde2024natural,
  title={Natural language supervision for general-purpose audio representations},
  author={Elizalde, Benjamin and Deshmukh, Soham and Wang, Huaming},
  booktitle={ICASSP},
  year={2024}
}

@inproceedings{cuturi2017soft,
  title={Soft-DTW: A differentiable loss function for time-series},
  author={Cuturi, Marco and Blondel, Mathieu},
  booktitle={ICML},
  year={2017}
}

@incollection{loper2023smpl,
  title={SMPL: A skinned multi-person linear model},
  author={Loper, Matthew and Mahmood, Naureen and Romero, Javier and Pons-Moll, Gerard and Black, Michael J},
  booktitle={Seminal Graphics Papers, Vol. 2},
  year={2023}
}

@inproceedings{siyao2022bailando,
  title={Bailando: 3D dance generation by actor-critic GPT with choreographic memory},
  author={Siyao, Li and Yu, Weijiang and Gu, Tianpei and Lin, Chunze and Wang, Quan and Qian, Chen and Loy, Chen Change and Liu, Ziwei},
  booktitle={CVPR},
  year={2022}
}

@inproceedings{law2009evaluation,
  title={Evaluation of algorithms using games: The case of music tagging},
  author={Law, Edith and West, Kris and Mandel, Michael I and Bay, Mert and Downie, J Stephen},
  booktitle={ISMIR},
  year={2009}
}

@inproceedings{engel2017neural,
  title={Neural audio synthesis of musical notes with WaveNet autoencoders},
  author={Engel, Jesse and Resnick, Cinjon and Roberts, Adam and Dieleman, Sander and Norouzi, Mohammad and Eck, Douglas and Simonyan, Karen},
  booktitle={ICML},
  year={2017}
}

@inproceedings{soleymani20131000,
  title={1000 songs for emotional analysis of music},
  author={Soleymani, Mohammad and Caro, Micheal N and Schmidt, Erik M and Sha, Cheng-Ya and Yang, Yi-Hsuan},
  booktitle={ACM Workshop on Crowdsourcing for Multimedia},
  year={2013}
}

@inproceedings{wang2025choreomuse,
  title={ChoreoMuse: Robust Music-to-Dance Video Generation with Style Transfer and Beat-Adherent Motion},
  author={Wang, Xuanchen and Wang, Heng and Cai, Weidong},
  booktitle={ACM MM},
  year={2025}
}

@inproceedings{baevski2020wav2vec,
  title={wav2vec 2.0: A framework for self-supervised learning of speech representations},
  author={Baevski, Alexei and Zhou, Yuhao and Mohamed, Abdelrahman and Auli, Michael},
  booktitle={NeurIPS},
  year={2020}
}

@article{oord2018representation,
  title={Representation learning with contrastive predictive coding},
  author={Oord, Aaron van den and Li, Yazhe and Vinyals, Oriol},
  journal={arXiv preprint arXiv:1807.03748},
  year={2018}
}

@article{rubner2000earth,
  title={The earth mover's distance as a metric for image retrieval},
  author={Rubner, Yossi and Tomasi, Carlo and Guibas, Leonidas J},
  journal={International Journal of Computer Vision},
  year={2000}
}

@article{tzanetakis2002musical,
  title={Musical genre classification of audio signals},
  author={Tzanetakis, George and Cook, Perry},
  journal={IEEE Trans. Speech Audio Process.},
  year={2002}
}

@article{dhariwal2020jukebox,
  title={Jukebox: A generative model for music},
  author={Dhariwal, Prafulla and Jun, Heewoo and Payne, Christine and Kim, Jong Wook and Radford, Alec and Sutskever, Ilya},
  journal={arXiv preprint arXiv:2005.00341},
  year={2020}
}

@article{loshchilov2017decoupled,
  title={Decoupled weight decay regularization},
  author={Loshchilov, Ilya and Hutter, Frank},
  journal={arXiv preprint arXiv:1711.05101},
  year={2017}
}

@article{davies2009evaluation,
  title={Evaluation methods for musical audio beat tracking algorithms},
  author={Davies, Matthew EP and Degara, Norberto and Plumbley, Mark D},
  journal={QMUL Tech. Rep. C4DM-TR-09-06},
  year={2009}
}

\end{document}